\documentclass[pre,aps,floats,superscriptaddress,floatfix]{revtex4}
\usepackage{amssymb,amsmath}
\usepackage{amsmath,amssymb}
\usepackage{graphicx}
\usepackage{psfrag}

\def\beq{\begin{equation}}
\def\eeq{\end{equation}}
\def\bea{\begin{eqnarray}}
\def\eea{\end{eqnarray}}
\begin{document}
\title{Active-to-absorbing state phase transition in the presence of fluctuating environments: Weak and strong dynamic scaling}
\author{Niladri Sarkar}\email{niladri.sarkar@saha.ac.in}
\author{Abhik Basu}\email{abhik.basu@saha.ac.in}
\affiliation{Theoretical Condensed Matter Physics Division, Saha
Institute of Nuclear Physics, Calcutta 700064, India}

\date{\today}

\begin{abstract}
We investigate the scaling properties of phase transitions between
survival and extinction (active-to-absorbing state phase transition,
AAPT) in a model, that by itself belongs to the directed percolation
(DP) universality class, interacting with a spatio-temporally
fluctuating environment having its own non-trivial dynamics. We
model the environment by (i) a randomly stirred fluid, governed by
the Navier-Stokes (NS) equation, and (ii) a fluctuating surface,
described either by the Kardar-Parisi-Zhang (KPZ) or the
Edward-Wilkinson  (EW) equations. We show, by using a one-loop
perturbative field theoretic set up, that depending upon the spatial
scaling of the variance of the external forces that drive the
environment (i.e., the NS, KPZ or EW equations), the system may show
{\em weak} or {\em strong dynamic scaling} at the critical point of
active to absorbing state phase transitions. In the former case AAPT
displays scaling belonging to the DP universality class, whereas in
the latter case the universal behavior is different.
\end{abstract}

\maketitle

\section{Introduction}


The simple epidemic process with recovery or the {\em Gribov
process} \cite{grib1,grib2} serves as an example of a prototypical
nonequilibrium phase transition. It, also formally known as the
Reggeon field theory \cite{rft1,rft2,rft3}, is a stochastic
multiparticle process that describes the essential features of local
growth processes of populations in a uniform environment near their
extinction threshold \cite{popdyn2,popdyn3} and belongs to the
Directed Percolation (DP) universality class.
%
 For recent reviews see
Refs.~\cite{hinrev,uwe-hans-review}. A simple realization of the DP
process is the predator-prey cellular automaton models. These models
display nonequilibrium active to absorbing state phase transitions
(AAPT) separating active from inactive or absorbing states due to
competitions between spontaneous particle decay (death) and particle
production (birth) processes. In these models for certain parameter
values, the steady state density is zero, i.e., the species gets
extinct. By definition, once reached an absorbing configuration, the
system can not escape from these configurations~\cite{marro}.
Experimental realizations of DP universality has been reported in
Ref.~\cite{chate} recently, where transitions between two
topologically different turbulent states of nematic liquid crystals
in their electrohydrodynamic convection regimes were observed.
Ref.~\cite{chate} measured the relevant scaling exponents with high
accuracy and found them to belong to the DP universality class. The
critical behavior of the AAPT \cite{hinrev} depends on the
conservation laws in the dynamics and the underlying symmetry. It
has been conjectured in the form of the {\em Directed Percolation
Hypothesis} \cite{popdyn3,dpuniv} that in the absence of any special
symmetry or conservation laws the AAPT belongs to the DP
universality class as long as the system has a
single absorbing state. 
In a more realistic situation, the spreading process in DP can also
be long ranged. For example, consider infecting agents being
advected by a local velocity field (e.g., parasites being carried in
a wind flow in an ecological system). 
To analyze such a situation, Ref.~\cite{grass20} introduced a
variation of the epidemic process with an infection probability
distribution that decays with the distance $r$ as a power law. Such
long range DP models have analyzed in details numerically
\cite{hin-how}, as well as analytically by using field theoretic
methods \cite{field1,field2}. In particular, Ref.~\cite{field2}
demonstrated four possible set of universal exponents for the long
range DP problem, corresponding to four different pairs of
(renormalization group) fixed point values for the two coupling
constants in the model, reflecting several possible universal
scaling behavior. Ref.~\cite{field2} enumerates the critical
exponents at each of the fixed points and their stabilities. More
recently, Ref.~\cite{antonov} discusses the effects of temporally
$\delta$-correlated multiplicative noises on the universal
properties of the AAPT. They discussed the condition under which the
usual DP universality class becomes unstable with respect to
perturbations by the multiplicative noises.

In the usual models belonging to the DP universality, the parameters
defining the models are taken to be constants. Thus they represent a
constant (non-fluctuating) environment. In this article we discuss
the general question of how interactions with a spatially and
temporally fluctuating environment affects the statistical
properties of a percolating agent or the density of a near-extinct
population $\phi$ at the critical point of an AAPT belonging to the
DP universality class, by coupling the environmental fluctuations
explicitly with the growth process. In order to address this
question, we consider two different cases of fluctuating
environments separately, namely, (i) a randomly stirred fluid,
described by the Navier-Stokes (NS) equation, and (ii) a fluctuating
surface, modeled by the Kardar-Parisi-Zhang (KPZ) or
Edward-Wilkinson (EW) equations.  We do not consider any feedback of
$\phi$ to the environment, i.e., the time evolution of the
environment is autonomous. In all the cases, the coupling with the
environment is nonlinear. Further we drive the environment (i.e.,
the NS, KPZ or EW equations) by Gaussian stochastic long-ranged
forces with specified variances. Because of our choice for the
dynamics of the environment, it itself displays universal spatial
and dynamical scaling in the long wavelength limit, with the
universality class depending upon the model used. In each of the
cases we calculate the relevant scaling exponents at the critical
point. The main result that we find is that the scaling behavior of
the system near the extinction point of the AAPT depends generally
upon the location of the system in the phase space spanned by
dimensionality $d$ and a parameter $y$ that characterizes the
spatial scaling of the noise correlations in the NS, KPZ or EW
models, and depending upon the location in the phase space, the
model may exhibit {\em strong dynamical scaling}, when the dynamics
of the field $\phi$ and the environment are characterized by the
same dynamic exponent, or {\em weak dynamical scaling}, when the two
have different dynamic exponents. Specifically in our model, weak
dynamic scaling implies scaling properties of $\phi$ belonging to
the DP universality class, where as strong dynamic scaling
corresponds to non-DP behavior. It may be noted that AAPT in
contact with a randomly stirred velocity field has been considered
in Ref.~\cite{antonov1}. We compare our results and scheme of
calculations with Ref.~\cite{antonov1} later.

Motivations of our work are {\em both theoretical and
phenomenological}. Our results are similar to those in
Ref.~\cite{field2}, and provides for a mechanism to introduce long
range flight in the otherwise local models of epidemic with
recovery.   Presence of both weak and strong dynamic scaling (in
different regions of the phase space) makes it a good candidate to
study general issues related to dynamical scaling in NESS. Apart
from theoretical motivations, the models and results discussed here
are useful in more biologically motivated context. Let us consider
the specific example of  a bacteria colony (or a biofilm of
bacteria) undergoing simple cell division  and death in its ordered
(nematic or polar) phase. The nematic or polar order parameter being
a broken symmetry mode obeys a dynamics that is scale invariant.
This should be coupled to the dynamics of the bacteria density,
which in addition to the growth decay terms, should have
symmetry-allowed couplings with the order parameter field. Moreover,
bacteria being living systems, there should be specific
nonequilibrium stresses (or {\em active stresses}) arising from the
continuous energy consumption of the living bacteria. Thus one
obtains a coupled model of a density undergoing the extinction
transition and a broken-symmetry order parameter field executing
scale invariant dynamics. Our results here will help us
understanding the interplay between the density field and the order
parameter field in determining the universal scaling near the
extinction threshold for more realistic naturally occurring systems.
The rest of the paper is organized as follows: In Sec.~\ref{review}
we discuss the basic DP model and the results from it in some
details. Then in Sec.~\ref{rand} we discuss the case when the
population density is coupled with a randomly stirred fluid
described by the NS equation. Within a one-loop DRG approach we
analyze the fixed points, calculate the associated scaling exponents
determined by the stable fixed points and present a linear stability
diagram. In Sec.~\ref{surf} we consider the environment to be a
fluctuating surface, which is governed either by the KPZ or the EW
equations. In all the cases we obtain the relevant scaling exponents
by one-loop renormalized perturbative calculations. In
Sec.~\ref{summary} we summarize and discuss our results.

\section{Equations of motion}
\subsection{Directed Percolation model}\label{review}

In order to set up the calculational background of our model, let us
briefly revisit the problem of extinction of a single species in a
uniform environment and  its universal properties near the
extinction threshold.  As an example, let us consider population
dynamics with population growth rate that depends linearly on the
local density of the species and death rate that depends
quadratically on the local density also undergoing a non-equilibrium
active to absorbing state (i.e., species extinction) phase
transition whose long distance large time properties are
well-described by the DP universality class. In terms of a local
particle number density $\phi({\bf x},t)$ and taking diffusive modes
into account, the Langevin equation that describes such a population
dynamics is given by [see, e.g., Ref.~\cite{uwe-hans-review}]
\begin{equation}
\frac{\partial\phi}{\partial t} = D\nabla^2\phi + \lambda_g\phi -
\lambda_d\phi^2 + \sqrt\phi \zeta,\label{dpeq}
\end{equation}
where $D$ is the diffusion coefficient, $\lambda_g$ is the growth
rate and $\lambda_d$ the decay rate of the density field $\phi$.
Stochastic function $\zeta({\bf x},t)$ is a zero-mean, Gaussian
distributed white noise with a variance given by
\begin{equation}
\langle\zeta({\bf x},t)\zeta(0,0)\rangle = 2D_2 \delta ({\bf
x})\delta (t). \label{vari1}
\end{equation}
The multiplicative nature of the effective noise in Eq.~(\ref{dpeq})
ensures the in-principle existence of an absorbing state ($\phi=0$)
in the system. Equation (\ref{dpeq}) allows us to extract the
characteristic length $\xi \sim \sqrt {D/|\lambda_g|}$ and diffusive
time scale $t_c\sim \xi^2/D\sim D/|\lambda_g|$ on dimensional
ground, both of which diverge upon approaching the critical point at
$\lambda_g=0$. Upon defining the critical exponents in the usual way
\cite{uwe-hans-review}
\begin{equation}
\langle \phi ({\bf x},t\rightarrow\infty)\rangle \sim
\lambda_g^\beta,\;\;\langle \phi({\bf x},t)\rangle \sim
t^{-\alpha}\; (\lambda_g=0),\;\; \xi\sim \lambda_g^{-\nu},\;\;
t_c\sim \xi^z_\phi/D\sim \lambda_g^{-z_\phi\nu}, \label{scaling}
\end{equation}
we identify the mean-field values
\begin{equation}
\beta=1,\alpha=1,\nu=1/2,\,\, {\mbox {and}},\,\, z_\phi=2.\label{mean}
\end{equation}
Further, the anomalous dimension $\eta$, which characterizes the
spatial scaling of the two-point correlation function, is zero
\cite{uwe-hans-review}. It is, however, well-known that fluctuations
are important near the critical point and as a result mean-field
values for the exponents are quantitatively inaccurate. In order to
account for the fluctuation effects, dynamic renormalization group
(DRG) calculations have been performed over an equivalent path
integral description of the Langevin
Eq.~(\ref{dpeq})~\cite{uwe-hans-review}. A one-loop renormalized
theory with systematic $\epsilon$-expansion, $\epsilon=d_c-d$, where
the upper critical dimension $d_c=4$ for this model, yields
\cite{uwe-hans-review},
\begin{equation}
z=2-\epsilon/12,\eta=\epsilon/12 \,\,\mbox{and}\,\, {1 \over \nu}=2
+ \epsilon/4. \label{dpexpo}
\end{equation}
 The DP universality class, characterized by the scaling exponents
 (\ref{dpexpo}) above,
is fairly robust, a feature formally known as the directed
percolation (DP) hypothesis~\cite{dpuniv}.  Only when one or more
conditions of the DP hypothesis are violated, one finds new
universal properties. For instance, the presence of long range
interactions are known to modify the scaling behavior:
Ref.~\cite{field2} examines the competition between short and long
ranged interactions, and identified four different possible phases.
Our results assume importance in this backdrop: In our model,
long-ranged interactions arise not because of any long range
hopping, but due to coupling of the density $\phi$ with an {\em
environment}, modeled by a randomly stirred fluid or a growing and
fluctuating surfaces, all of which in turn are driven by long range
noises. It is expected that the presence of these long-range
correlated background may affect the universal scaling behavior; our
results below confirm this in general.

\subsection{Advection of simple epidemic process by a randomly
stirred fluid} \label{rand}

Here we consider a randomly stirred fluid as a fluctuating
environment coupled to the AAPT of a population density near its
extinction threshold.

\subsubsection{Randomly stirred fluid model} \label{rand1}

The dynamics of the fluid in the incompressible limit is described
by a velocity field $\bf v$ which follows the Navier-Stokes equation
\cite{landau}
 \bea
\partial_t v_i + \lambda_1({\bf v}\cdot\nabla)v_i = D_\nu\nabla^2 v_i - \frac{\nabla_i p}{\rho} + f_i,
\label{ns} \eea
 together with the incompressibility condition $\nabla\cdot {\bf
 v}=0$, which may be used to eliminate pressure in the usual way.
 Here  $p$ the pressure, $\rho$ the
 density (a constant in the incompressible limit) and $D_\nu$ the
 kinematic viscosity. Parameter $\lambda_1$ is a non-linear coupling
 constant, which does not renormalize in the hydrodynamic limit under mode elimination due to the Galilean invariance of  Eq.~(\ref{ns}) \cite{fns,yakhot}.
 Function
$\bf f$ is the external
 force required to maintain a driven steady state. A promising
starting point for theoretical/analytical studies on homogeneous and
isotropic externally stirred fluid  is the randomly forced
Navier-Stokes model \cite{fns,yakhot}, where  $\bf f$ is assumed to
be a stochastic function which is  zero-mean and Gaussian
distributed with a
 variance
 \cite{yakhot}
\begin{equation}
\langle f_i({\bf q},t)f_j({\bf -q}, 0)\rangle=\delta_{ij}{2D_1\over
q^{d-4+y}}\delta (t) \label{yakhotnoise}
\end{equation}
 in $d$-dimensions, $\bf q$ is a Fourier wavevector, $D_1$ is a constant amplitude and $y>0$. We are interested when $d-4+y>0$. Since in this case
 the variance (\ref{yakhotnoise}) diverges in the hydrodynamic
 limit, the force $\bf f$ is said to be long-ranged and is infra red (IR) singular. This is a variant of the Model B
(with the identification $y=2-d$) of  Ref.~\cite{fns} which was
subsequently used in Ref.~\cite{yakhot} to calculate scaling
behavior and various universal quantities associated with
homogeneous and isotropic $3d$ fluid turbulence. The velocity field
shows universal spatial and dynamical scaling independent of the
microscopic viscosity and forcing amplitude\cite{foot}. These
scaling properties are characterized by the roughness exponent
$\chi_v$ and dynamic exponent $z_E$ defined through the definition
 \beq
 \langle v_i ({\bf x},t)v_i(0,0)\rangle\sim |{
 x}|^{2\chi_v}f_v(|x|^{z_E}/t),\label{scalev}
  \eeq
  where $f_v$ is a scaling function. Invariance under the Galilean transformation
  (see below) yields that the coupling constant does not receive
  any fluctuation correction in the hydrodynamic limit. A
  consequence of that is the exact relation between the exponents
  $\chi_v$ and $z_E$ \cite{yakhot,jkb}:
$\chi_v + z_E=1$.
  Applications of one-loop
DRG to systems with long range noises are more complicated and less
controlled than its application in problems of equilibrium critical dynamics
 \cite{yakhot,jkb,ronis}. Despite the limitations, such
calculations are successful in obtaining several useful results on
dimensionless numbers and scaling exponents. Due to the infra red
singular nature of the bare noise variance (\ref{yakhotnoise}), the
perturbation theory does not generate any fluctuation correction to
the bare noise variance (\ref{yakhotnoise}) which is more singular
than it. Consequently, it is assumed to be unrenormalized. Thus in a
renormalized perturbation theory, only kinematic viscosity $D_v$
undergoes nontrivial renormalization. An explicit
  one-loop DRG calculation yields
  \beq
  z_E=2-y/3.\label{ze}
 \eeq
Further, $\chi_v=-1+y/3$. Thus for sufficiently high value of $y$,
$z_E$ may even be less than unity, resulting into {\em turbulent
diffusion} which is a very efficient way of mixing. In particular,
the value $y=d$ in (\ref{yakhotnoise}) is of particular physical
interest, since it corresponds to the famous K41 energy spectrum for
the velocity field: One finds for the one-dimensional energy
spectrum $E_v (q)\sim k^{d-1} \langle |v_i({\bf q},t)|^2\rangle \sim
q^{-5/3}$ in three dimensions.

\subsubsection{Extinction transition in contact with a randomly
stirred fluid} \label{rand2}

 In the standard models for epidemic with recovery belonging to the DP universality class, the population density field $\phi$ is allowed only
 to diffuse (apart from local reproduction and death). However in general, such processes (of reproduction and death) may take place in
 a fluid environment and the local population may get advected in addition to diffusing.  Thus they may
 be called
 {\em reaction-advection-diffusion} systems. For
 instance, a reaction or  birth/death of bacteria may
 take place in fluid, which in turn may be thermally fluctuating, or
 may be externally stirred. Here, we find out how the
 universal properties of
 standard AAPT are affected
 when the system is advected by an externally stirred fluid.

 The percolating field
$\phi$ satisfies the same equation (\ref{dpeq}), now supplemented by an
advective non-linearity:
\begin{equation}
\frac{\partial\phi}{\partial t} + \lambda_2 {\bf v}\cdot \nabla\phi=
D\nabla^2\phi + \lambda_g\phi - \lambda_d\phi^2 + \sqrt\phi
\zeta,\label{dpeqI}
\end{equation}
Here, $\lambda_2$ is a coupling constant, which advectively couples
$\phi$ with $\bf v$. Constants $\lambda_g$ and $\lambda_d$ denote
growth and decay rates.  Gaussian noise $\zeta$ has the same
variance as (\ref{vari1}). Redefining coefficients $\lambda_g=D\tau$
and $\lambda_d=\frac{Dg_2}{2}$ for calculational convenience, Eq.
(\ref{dpeqI}) may be written as
\begin{equation}
\frac{\partial \phi }{\partial t} + \lambda_2 {\bf v}\cdot \nabla
\phi=  D(\nabla^2+ \tau)\phi - \frac{Dg_2}{2}\phi^2+ \sqrt\phi
\zeta. \label{dpeqF}
\end{equation}
 Thus the critical point is now defined by (renormalized) $\tau=0$.

As for the usual DP problem, the system exhibits a continuous phase
transition from active to absorbing states as (renormalized or
effective) $\tau\rightarrow 0$. The associated universal scaling
exponents are formally defined as in (\ref{scaling}) above. At the
mean-field level the model Eqs.~(\ref{ns}) and (\ref{dpeqF}) yield
the same mean-field values for the scaling exponents as above in
Sec.\ref{review}. Nonlinear couplings $\lambda_2$ and $Dg_2$, together
with the expected large fluctuations near the critical point and
multiplicative nature of the long ranged noise with variance
(\ref{yakhotnoise}) are expected to substantially alter the
mean-field values of the exponents. Studies of these in-principle
require full solution for the field $\phi({\bf x},t)$. Equations
(\ref{ns}) and (\ref{dpeqF}) being nonlinear, cannot be solved
exactly. Hence perturbative means are necessary. We address this
issue systematically via
 standard implementation of DRG procedure, based on a one-loop perturbative expansion
 in the  coupling constants  $\lambda_2$ and $Dg_2$ about the linear theory. The resulting perturbative
 corrections to the correlation
function may be equivalently viewed as arising from modifications
(renormalization) of the parameters and fields in the model
Eqs.~(\ref{ns}) and (\ref{dpeqF}).

We being with the Janssen-De Dominicis dynamic generating functional
\cite{bausch} corresponding the Langevin equations (\ref{ns}) and
(\ref{dpeqF}) together with the corresponding noise variances
(\ref{vari1}) and (\ref{yakhotnoise}), which is given by
 \bea {\mathcal Z}_{NS}=\int Dv
D\hat{v} D\phi D\hat{\phi} \exp[S_{NS}],
 \eea
  where $\hat{\phi}$ and
$\hat{v}$ are auxiliary fields corresponding to the dynamical fields
$\phi$ and $\bf v$ respectively which appear due to elimination of
the noises from the generating functional $\mathcal{Z}_{NS}$. The
action functional $S_{NS}$ is written as
 \bea
  S_{NS}({\bf v},{\bf \hat{v}}, \phi, \hat{\phi}) &=&
D_1\int \frac{d^dk}{(2\pi)^d}\int dt\, \hat{v}_i\hat{v}_j
P_{ij}({\bf k})k^{4-y -d}
- \int\frac{d^dk}{(2\pi)^d}\int dt\,\hat{v}_i\{ \partial_t v_i -
i\frac{\lambda_1}{2}P_{ijl}({\bf k})\sum_{\bf q} v_j({\bf q})v_l({\bf k-q}) + D_\nu k^2v_i\} \nonumber \\
- \int\frac{d^dk}{(2\pi)^d}\int dt\,\hat{\phi}\{ \partial_t\phi &-&
i\lambda_2k_l\sum_{\bf q}v_l({\bf q})\phi({\bf k-q}) + D (k^2 -
\tau)\phi + {D g_2 \over 2}\sum_{\bf q}\phi({\bf q})\phi({\bf k-q})
- \frac{Dg_1}{2} \sum_{\bf q}\hat{\phi}({\bf q})\phi({\bf k-q})\},
 \label{action1}
  \eea
 where $D_2=\frac{Dg_1}{2}$, $P_{ij}({\bf k})=\delta_{ij}-{k_ik_j \over k^2}$
is the transverse momentum operator and
$P_{ijl}({\bf k})=P_{ij}({\bf k})k_l+ P_{il}({\bf k})k_j$. Note that the last
two non-linear terms in (\ref{action1}) {\em do not} have the same
coupling constant $Dg/2$, unlike the usual DP problem. This is
consistent with the lack of invariance of $S_{NS}$ under the {\em
rapidity symmetry}. The rapidity symmetry of the original DP problem
(see, e.g., \cite{uwe-hans-review}) is no longer  admissible in the
present case, since the Navier-Stokes Eq.~(\ref{ns}) being a viscous
dissipative equation cannot be invariant under time inversion.

Before we present our detailed DRG calculation, let us note the
following: Since the dynamics of the randomly stirred fluid is
independent of $\phi$, its dynamic exponent $z_E$ should be same as
that in the absence of any percolating agent. We have $z_E=2-y/3$
(see discussions above and the calculations below). Thus if there is
a regime characterized by {\em strong dynamic scaling}, we should
have $z_\phi=z_E=2-y/3$. In this regime, the nonlinearity of the
basic DP process (i.e., coupling constants $g_1$ and $g_2$) may or
may not be relevant in an RG sense, corresponding to what we call
LDP (long-range DP) and LR (long range) phases having different
static scaling properties (the dynamic exponents are same in LDP and
LR phases), characterized by the LDP and LR fixed points (FP),
respectively. In contrast, when the coupling constant $\lambda_2$
that couples $\phi$ and $\bf v$ is irrelevant in DRG sense, the
dynamics of $\phi$ is independent of $\bf v$ and hence $\phi$
displays a dynamics that is identical to the usual DP problem with a
dynamic exponent $z_\phi=z_{DP}=2-\epsilon/12$. Can there be a phase
displaying weak dynamic scaling with $\lambda_2$ still being
relevant? We expect not; because if $\lambda_2$ is indeed relevant
(in an RG sense), the the whole action (\ref{action1}) including the
coupling term $\hat\phi {\bf v}\cdot\nabla\phi$ must be invariant
under combined rescaling of space, time and fields characterized by
a single set of exponents, i.e., a single dynamic exponent. Thus the
assumption of weak dynamic scaling, i.e., the existence of two
unequal dynamic exponents, rules out the DRG relevance of
$\lambda_2$. Finally one may in principle have a phase where {\em
all} nonlinerities are irrelevant with the $\phi$-field being
characterized by the exponents of the linear theory (Gaussian phase)
which turns out to be always linearly unstable. Thus, in short, we
expect four different phases characterized by four FPs (LR, LDP, DP
and Gaussian) and their associated set of exponents. Some of these
phases may not be linearly stable. We shall confirm these physically
inspired picture through detailed one-loop calculations below. In
addition, we calculate all the critical exponents as well.

Equations (\ref{ns}) and (\ref{dpeqI}) are invariant under the
Galilean invariance
 \beq
 {\bf v}\rightarrow {\bf v}+ {\bf u}_0,\,\,\phi\rightarrow\phi,\,\,{\bf x}\rightarrow {\bf x}+
 {\bf u}_0t,\,\,t\rightarrow t,\,\,\frac{\partial}{\partial
 t}\rightarrow \frac{\partial}{\partial t}. \label{galns}
 \eeq
Invariance of the system under (\ref{galns}) ensures that the
coupling constants $\lambda_1$ and $\lambda_2$ are equal and do not
renormalize in the hydrodynamic limit: We henceforth set
$\lambda_1=\lambda_2=\lambda$ below. The role of the (bare or {\em
unrenormalized}) coupling constants in the ordinary perturbation
theory in the present model is played by
$u=g_1g_2$ and $w=\frac{\lambda^2 D_1}{D_\nu^3}$.
 In addition, there is a dimensionless number
$\theta=D/D_\nu$, the {\em Schmidt number} which characterizes the
ensuing NESS of the AAPT, and is a control parameter of the model.
We set up a renormalized perturbative expansion in $\epsilon=4-d$
and $y$ up to the one-loop order.  In order to ensure ultra-violet
(UV) renormalization of the present model, we are required to render
finite all the non-vanishing two- and three-point functions by
introducing multiplicative renormalization constants. This procedure
is well-documented in the literature, see, e.g.,
Ref.~\cite{uwe-book}. Here, vertex functions of different orders are
formally defined by appropriate functional derivatives of
$\Gamma[{\bf v},{\bf \hat v},\phi,\hat\phi]$ with respect to various
fields (dynamical and auxiliary), where $\Gamma$ is the vertex
generating functional and the Legendre transform of $\ln Z_{NS}$
\cite{uwe-book}.  In the present model the following vertex
functions show primitive divergence at the one-loop level:
$ (i)\frac{\delta^3 \Gamma}{\delta v_i \delta \hat
 v_i},\,(ii)\frac{\delta^2\Gamma}{\delta\phi\delta\hat\phi},\,(iii)
 \frac{\delta^3\Gamma}{\delta\hat\phi\delta\phi\delta\phi},\,(iv)
 \frac{\delta^3\Gamma}{\delta\hat\phi\delta\hat\phi\delta\phi}.$
 Their divergences may be absorbed by introducing renormalization $Z$-factors
 (see below).

 We employ the dimensional regularization scheme to compute the
momentum integrals associated with the one-loop vertex function
renormalization, and choose $\tau=\mu^2$ as our normalization point,
where $\mu$ is an intrinsic momentum scale of the renormalized
theory. The scaling behavior of the correlation or vertex functions
in the hydrodynamic limit may be extracted by finding their
dependence on $\mu$ by using the renormalization group (RG)
equation, which may in turn be obtained from the one-loop
renormalization $Z$-factors. The Galilean invariance of the present
model leads to exact Ward identities between certain two- and
three-point vertex functions, which yields that the coupling
constant $\lambda$ does not renormalize \cite{foot2}.  Further,
since the bare noise variance (\ref{yakhotnoise}) is IR-singular,
perturbation theories do not generate any correction to it which is
more singular than it. Thus, the coefficient $D_1$ also does not
renormalize. Fields $\bf v$ and $\bf \hat v$ do not re
normalize as
well.
The renormalized fields and parameters (denoted by a superscript
$R$) are defined through the corresponding $Z$-factors as
 \bea \phi=Z\phi^R \,\,, \,\, \hat{\phi}=\hat{Z}\hat{\phi}^R,\;\;\tau=Z_\tau\tau^R \,\, ,
\,\, D=Z_D D^R \,\,, \,\, g_1=Z_{g_1}g_1^R\,\,,\,\,g_2=Z_{g_2}g_2^R
\,\,,\,\, D_\nu=Z_{D_\nu}D_\nu^R. \label{zpara}
 \eea
We perform explicit one-loop calculations in terms of coupling
constants $u=g_1g_2$ and $w=\lambda^2 D_1/D_\nu^3$. From
definition we have $Z_u=Z_{g_1}Z_{g_2}$ and $Z_w=Z_{D_\nu}^{-3}$.
Now absorbing factors of $1/16\pi^2$ into these coupling constants
i.e, $u\rightarrow u/16\pi^2$ and $w\rightarrow w/16\pi^2$ we can
write down the Z-factors in terms of these scaled coupling
constants. Further, since one of the $Z$-factors from the set
$Z,\hat Z, Z_{g_1},Z_{g_2},Z_D,Z_\tau,Z_{D_\nu}$ is redundant, we use this
freedom to set $Z=\hat Z$ without any loss of generality. We obtain
\begin{eqnarray}
Z_{D_\nu} &=& 1-{aw \over y}\mu^{-y}, \\
Z&=&\hat Z=1+\frac{u}{8}\frac{\mu^{-\epsilon}}{\epsilon},\\
Z_D&=&1-\frac{u}{8}\frac{\mu^{-\epsilon}}{\epsilon}-\frac{\lambda^2
D_1}{D_\nu D(D_\nu +D)}(1-
\frac{1}{d})\frac{2}{16\pi^2}\frac{\mu^{-y}}{y},\\
Z_\tau&=&1+\frac{\lambda^2 D_1}{D_\nu D(D_\nu
+D)}(1-\frac{1}{d})\frac{2}{16\pi^2}\frac{\mu^{-y}}{y}-\frac{u}{8}\frac{\mu^{-\epsilon}}{\epsilon},\\
Z_u&=&1+\frac{3u}{2}\frac{\mu^{-\epsilon}}{\epsilon} +
4\frac{\lambda^2 D_1}{D_\nu D(D_\nu
+D)}(1-\frac{1}{d})\frac{\mu^{-y}}{y}.\label{zfacns}
\end{eqnarray}
We find from the explicit one-loop results (\ref{zfacns}) that the
$Z$-factors  are linear in $u$, $w$ and ${\tilde {w}}=\lambda^2
D_1/[D_\nu D (D_\nu + D)]$,  (but not in $\theta$ directly). Thus
the perturbative expansions are in effect expansions in powers of
$u$, $w^R$ and $\tilde w$. Further, $Z_\theta = Z_D / Z_{D_\nu}$ is
linear in $\tilde w$, but not in $\theta$ itself. In contrast
$Z_{\tilde w}$ cannot be expressed as a linear function of $\tilde w
$, along with $u$. In order to simplify the calculation, we make an
approximation that $\theta\ll 1$, such that $\tilde w=1/(D_\nu^2
D)$. This allows us to write $Z_{\tilde w}$ linearly in terms of
$\tilde w$. We show that, despite our simplifying assumption, we are
able to obtain physically meaningful and interesting results which
we present below. In the limit of small $\theta$ we get
$Z_{\tilde{w}} =Z_wZ_{D_\nu}/Z_D$. Using $Z_w=Z_{D_\nu}^{-3}$, we
can write $Z_{\tilde{w}} =Z_{D_\nu}^{-2}Z_D^{-1}$. Thus to derive
$Z_{\tilde{w}}$ we should find out what $Z_D$ is $\theta\ll 1$. It
turns out to be (after absorbing a factor of $1/16\pi^2$ in the
definition of $\tilde w$, setting $d=4-\epsilon$ and keeping the
lowest order term in $\epsilon$)
 \bea Z_D &=& 1-{u \over
8\epsilon}\mu^{-\epsilon} - {3\tilde{w} \over 4y}\mu^{-y}, \eea and
hence \bea Z_{\tilde{w}}=1+{u \over 8\epsilon}\mu^{-\epsilon} + {2w
\over y}\mu^{-y} + {3\tilde{w} \over 4y}\mu^{-y}, \eea
  where $a=\frac{3}{2}\frac{d-1}{d-2}$.
  The Wilson's flow functions can be
defined as \bea \zeta_\phi &=&\mu{\partial \over \partial\mu}\ln
Z,\,\, \zeta_{\hat\phi}=\mu{\partial \over \partial\mu}\ln \hat{Z},
\,\, \zeta_D = \mu{\partial \over \partial\mu}\ln Z_D, \,\,
\zeta_\tau = \mu{\partial \over \partial\mu}\ln Z_\tau -2, \,\,
\zeta_{D_\nu} = \mu{\partial \over \partial\mu}\ln Z_{D_\nu},
\label{zetadnu} \eea
 and the $\beta$-functions as
\bea \beta_u = \mu{\partial \over \partial\mu} u^R \,\,,\,\, \beta_w
= \mu{\partial \over \partial\mu} w^R \,\,,\,\, \beta_{\tilde{w}} =
\mu{\partial \over \partial\mu}\tilde{w}^R. \eea
The renormalized coupling constants are written as 
$u^R=uZ_u^{-1}\mu^{-\epsilon} \,\,,\,\, w^R=wZ_w^{-1}\mu^{-y}
\,\,,\,\, \tilde{w}^R=\tilde{w}Z_{\tilde{w}}^{-1}\mu^{-y}$,
which
 for the $\beta$-functions yields
 \bea \beta_w =
w^R(-y + 3aw^R) \,\,,\,\, \beta_u = u^R\left( -\epsilon + {3 \over
2}u^R + {3 \over 2}\tilde{w}^R \right) \,\,,\,\,
\beta_{\tilde{w}}=\tilde{w}^R \left( -y + 2w^R + {3 \over
4}\tilde{w}^R + {1 \over 8}u^R\right). \eea Fixed points of the
model are given by the solutions of
$\beta_u=\beta_w=\beta_{\tilde{w}}=0$ which gives us various
solutions depending on the different values of the the parameters
$u^R$, $w^R$ and $\tilde{w}^R$. The only stable fixed point solution
for $w_R$ is $w_R= y/(3a)$. The different fixed point solutions for
$u_R$ and $\tilde w_R$ are as follows:
 \bea
a &\rangle& \,\, u^R=0,  \tilde{w}^R=0 \,\,\, \mbox{ (the trivial Gaussian fixed point)} \\
b &\rangle& \,\, u^R={2 \over 3}\epsilon,  \tilde{w}^R=0 \,\,\,
\mbox{(the  DP fixed point)} \\
c &\rangle& \,\, u^R=0,  \tilde{w}^R={4 \over 9}y \,\,\, \mbox{(LR, long range fixed point)}\\
d &\rangle& \,\, u^R={4 \over 5}\epsilon - {8 \over 15}y,
\tilde{w}^R= -{2 \over 15}\epsilon + {8 \over 15}y \,\,\,
\mbox{(LDP, long range DP fixed point)}.\label{ntfp} \eea
 Since by
construction $u^R$ and $\tilde w^R$ cannot be negative, from
Eq.~(\ref{ntfp}) it is obvious that the non-trivial fixed points may
be present in the range $4 y\geq\epsilon\geq {2 \over 3}y$. For
$\epsilon<{2 \over 3}y$ the system has fixed points $u^R=0$ and
$\tilde{w}^R={4 \over 9}y$ which defines the LR fixed point. For
$\epsilon>4y$ the system has fixed points $u^R={2 \over 3}\epsilon$
and $\tilde{w}^R=0$ which defines the DP fixed point. We will see
below that these ranges precisely coincide with the region of
stability of those fixed points.

The critical exponents are formally related to the Wilson flow
functions~\cite{uwe-hans-review} and hence to $u^R,\,w^R$ and
$\tilde w^R$, and are given by \bea \eta_\phi &=& \eta_{\hat{\phi}}
= -\zeta_{\phi} = {u^R \over 8},
 \\
 z_E&=&2-\zeta_{D_\nu}=2-aw^R,\\
z_\phi &=& 2-\zeta_D = 2-{u^R \over 8} - {3\tilde{w}^R \over 4},\\
\nu &=& -{1 \over \zeta_\tau}= [2+{3u^R \over 8} + {3\tilde{w}^R
\over 4}]^{-1}. \eea
 Thus the different exponents at different regions, defined by $u^R$ and $\tilde w^R$, of
 the phase space spanned by $\epsilon$ and $y$ are
 \begin{itemize}
 \item
 At the DP fixed point $({2\epsilon \over
3},0)$, the critical exponents are \bea
\eta_\phi=\eta_{\hat{\phi}}={\epsilon \over 12}\,\,,\,\,
z_\phi=2-{\epsilon \over 12},\,\,\,z_E=2-y/3, \,\,,\,\, {1 \over
\nu}= 2+{\epsilon \over 4}. \eea
 \item At the LR fixed point $(0,{4y \over
9})$, the critical exponents turn out to be
 \bea \eta_\phi=\eta_{\hat{\phi}}=0\,\,,\,\,
z_\phi=z_E=2-{y \over 3} \,\,,\,\, {1 \over \nu}= 2+{y \over 3}.
 \eea
 \item  At the LDP fixed point $({4\epsilon \over 5}-{8y
\over 15}, -{2\epsilon \over 15}+{8y \over 15})$, the critical
exponents are \bea \eta_\phi=\eta_{\hat{\phi}}={\epsilon \over
10}-{y \over 15}\,\,,\,\, z_\phi=z_E=2-{y \over 3} \,\,,\,\, {1
\over \nu}= 2+{\epsilon \over 5}+{y \over 5}. \eea
\end{itemize}
The interesting point to be noted here is that the dynamic exponent
$z_\phi$ has the same value for both LR and LDP FPs, where has the
corresponding static scaling exponents pick up different values. For
example, at the LR FP the anomalous dimensions $\eta_\phi$ and
$\eta_{\hat\phi}$, which describe the spatial scaling of the
correlation function and the propagator, are zero; hence spatial
scaling of the correlation function and the propagator are given by
the mean-field analysis (\ref{mean}), where as they pick non-trivial
fluctuation corrections at the LDP FP. See Sec.~\ref{summary} for
more discussions on this.

\begin{figure}[htb]
\includegraphics[height=5cm]{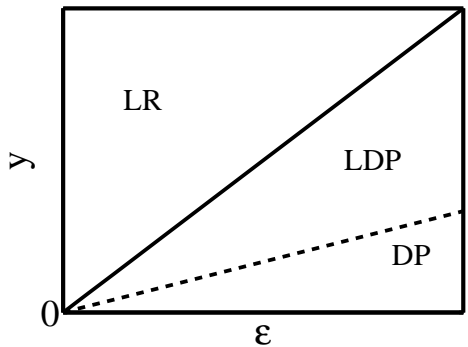}
\caption{Schematic phase diagram depicting the stable phases of the
model with a randomly stirred fluid as the environment. The
continuous line which is given by $4y=\epsilon$ is the boundary between
the LR and the LDP phases. The dashed line which is given by
$\epsilon=2y/3$ is boundary between the LDP and the DP phases.}
\label{nsfig}
\end{figure}

To analyse the stability of the fixed points we must evaluate the
matrix \bea M_{NS}= \left(\begin{array}{cc}
{\partial\beta_u \over \partial u} & {\partial\beta_u \over \partial \tilde{w}} \\
{\partial\beta_{\tilde{w}} \over \partial u} &
{\partial\beta_{\tilde{w}} \over \partial \tilde{w}}
\end{array}
\right) \label{stab}
 \eea
  and determine its eigenvalues $\Lambda$ at $u=u^R$ and $\tilde w=\tilde w^R$. The
condition for infrared stable fixed points is that all the
eigenvalues of the stability matrix $M_{NS}$ should be positive:
\begin{itemize}
\item At the Gaussian fixed point $(u^R,\tilde{w}^R)=(0,0)$ the
eigenvalues are $\Lambda=-\epsilon, -{y \over 3}$. As both the
eigenvalues are negative the fixed point is unstable.
 \item At
the fixed point $(0,{4 \over 9}y)$ the eigenvalues are
$\Lambda=-\epsilon + {2 \over 3}y, {1 \over 3}y$. The stability of
this fixed point depends on the condition $\epsilon\leq {2 \over
3}y$.
 \item For the DP fixed point $({2 \over 3}\epsilon,0)$ the
eigenvalues are $\Lambda=\epsilon, -{1 \over 3}y +{1 \over
12}\epsilon$. It is stable if $\epsilon\geq 4y$.
 \item For the fixed point $u^R = 4\epsilon/5 - 8y/15,\,\tilde w^R =
 8y/15 - 2\epsilon/15$, eigenvalues are
 $$\Lambda=\left[\frac{11\epsilon}{10} - \frac{2y}{5}\pm
 \left(\frac{131\epsilon^2}{100}+\frac{32y^2}{75}-\frac{101y\epsilon}{75}\right)^{1/2}\right]/2=\Lambda_\pm$$
 (say). Both eigenvalues are real. One of them $\Lambda_+$ is always
 positive within the window $4y>\epsilon> 2y/3$ (this is the window in which this fixed point exists). Since
 $\mbox{det}
 M_{NS}=2(3\epsilon -2y)(y-\epsilon/4)/15$ is negative outside this
 window, the second eigenvalue $\Lambda_-$ must change sign when
 $4y=\epsilon$ or $\epsilon=2y/3$. Thus this fixed point is stable
 for $4y>\epsilon>2y/3$.
\end{itemize}
We see that the lines demarcating the regions of stability between
the fixed points DP and LDP is given by $\tilde w^R=0$,
 which in turn is the same line where the
values of the dynamic exponents $z_\phi=2-\epsilon/12$ and
$z_\phi=2-y/3$, corresponding respectively to the the DP and LDP
fixed points, are equal, where as the line demarcating the regions
of stability between the fixed points LDP and LR are determined by
$u^R=0$. Further, as the system crosses over from the DP to LDP
fixed point, $z_\phi$ changes smoothly. The same is true for the
anomalous dimension $\eta_\phi=\eta_{\hat\phi}$, which smoothly
crosses over from its value $\epsilon/12$ at the DP fixed point to
$\epsilon/10 -y/15$ at the LDP fixed point to 0 at the LR fixed
point. The correlation length exponent $\nu$ shows similar behavior.
In Fig.~(\ref{nsfig}) below  a phase diagram of the stable phases in
the $\epsilon-y$ plane is shown.
Let us now compare with Ref.~\cite{antonov1} where AAPT in
contact with a randomly stirred fluid described the NS Eq. with a
long-ranged force is considered within a one-loop renormalized
perturbation theory like above. Our results for the scaling
exponents and the phase diagram are same as that in
Ref.~\cite{antonov1}. The differences between Ref.~\cite{antonov1}
and ours lie essentially in the details: Our choice for the coupling
constants is slightly different from Ref.~\cite{antonov1}. Let us
reconsider our choice for the effective coupling constants as used
in the calculations above: The expressions of the $Z$-factors in
(\ref{zfacns}) as well as $Z_\theta$ reveal that $u,\,w$ and $\tilde
w=\lambda^2 D_1/[D_\nu D(D_\nu +D)]$ appear as the bare
(dimensionless) expansion parameters in which the one-loop
perturbative expansions are linear; but these are not linear in
$\theta$. Finally, our assumption of (bare) $\theta \ll 1$ allows us
to write $Z_{\tilde w}$ as a linear function of $\tilde w$ as well.
In contrast to us, Ref.~\cite{antonov1} worked with $u,w$ and $e$
($\theta$ in our notation) as coupling constants in which the
perturbative expansions are made.  Our main operational motivation
of expanding in terms of $u,w$ and $\tilde w$ is that $u$ and
$\tilde w$ (or rather their renormalized counterparts $u_R$ and
$\tilde w_R$) directly describe the relative importance of the
original DP nonlinearity {\em vis-a-vis} the advective nonlinearity.
Hence, the plausibility of four possible phases described by
($u_R=0=\tilde w_R$), ($u_R\neq 0, \tilde w_R=0$), ($u_R\neq
0,\,\tilde w_R\neq 0$) and ($u_R=0,\,\tilde w_R\neq 0$) becomes
immediately clear, a fact borne out by the detailed calculations
described above. Nevertheless, there is no real contradiction
between this work and Ref.~\cite{antonov1} as is evident by the same
values for the scaling exponents and same phase diagram.

\subsection{Extinction transition in an environment of a fluctuating
surface}\label{surf}

 Consider next
birth/death process of an agent taking place on a fluctuating
surface: We thus now discuss the situation, where the dynamics of a
population density field $\phi$ near its extinction transition is
assumed to be in contact with  a fluctuating/growing surface without
any overhangs, represented by a scalar height field $h(x,t)$,
measured from an arbitrary substrate, which satisfies the nonlinear
KPZ equation \cite{kpzref} or the linear EW \cite{barabasi-book}
equation of motion.

\subsubsection{Edward-Wilkinson surface growth
equation}\label{ewrev}

The EW Eq.\cite{barabasi-book} is the simplest equation that
describes a growing surface by a single valued height field $h({\bf
x},t)$:
\begin{equation}
\frac{\partial h}{\partial t} = D_h \nabla^2 h + \psi.\label{ew}
\end{equation}
We consider the noise $\psi$ is zero-mean, Gaussian distributed with
a variance given by Eq.~(\ref{kpznoise}). We are concerned here with
the effects of long ranges noises. The variance of the noise
$\psi({\bf k},t)$ in the Fourier space is chosen to be
\begin{equation}
\langle \psi({\bf q},t)\psi (-{\bf q},0)\rangle = 2D_1
q^{2-y-d}\delta (t),\label{kpznoise}
\end{equation}
where $D_1$ is a constant setting the amplitude and the parameter
$y>0$. EW Eq.~(\ref{ew}) is {\em not} invariant under the Galilean
invariance. Eq.~(\ref{ew}), owing to its linearity, can be solved
exactly. In particular, its dynamic exponent $z_h=2$, regardless of
the value of $y$.

\subsubsection{Kardar-Parisi-Zhang equation} \label{kpzrev}

The KPZ equation \cite{kpzref} is a nonlinear generalization of the
EW Eq. above and is given by
\begin{equation}
\frac{\partial h}{\partial t} + \frac{\lambda_1}{2}(\nabla h)^2 =
D_h \nabla^2 h + \psi,\label{kpz}
\end{equation}
where $h$ is the height field which gives the height of the growing
surface from a reference plane, $\lambda_1$ a coupling constant,
$D_h$ is a diffusion constant, and $\psi$ is the external noise.
Evidently, the EW Eq.~(\ref{ew}) can be obtained from the KPZ
Eq.~(\ref{kpz}) by setting $\lambda_1=0$. Due to the Galilean (tilt)
invariance (see below) of the KPZ equation  the coupling constant
$\lambda_1$ does not renormalize. Analogous  to the NS
Eq.~(\ref{ns}) one defines dynamic exponent $z_h$ and the roughness
exponent $\chi_h$ for characterization of the correlations of the
fluctuations of $h$: One writes
\begin{equation}
\langle h({\bf x},t) h(0,0)\rangle \sim |x|^{2\chi_h}f_h (x^{z_E}/t),
\end{equation}
where $f_h$ is a scaling functions. Nonrenormalization of
$\lambda_1$ yields an exact relation $\chi_h+z_E=2$.
 When $\psi$ is a zero-mean Gaussian
distributed white noise, the KPZ equation describes a smooth to
rough phase transition at $d>2$ \cite{natterman}. We will however be
concerned here with the situation when the KPZ equation (\ref{kpz})
is driven by a long range noise, same as (\ref{kpznoise}).
Stochastic dynamics of the KPZ equation driven by a long ranged
correlated noise has already been studied extensively in
Ref.~\cite{medina} by using DRG methods. Such applications suffer
from several technical complications which are similar in nature to
those for the NS Eq.~(\ref{ns}) with a long-ranged noise.
Nevertheless, one-loop DRG calculations have been successful is
obtaining the scaling exponents. As for the randomly stirred fluid
model, the only quantity that renormalizes here is the diffusion
coefficient $D_h$. Explicit one-loop RG calculation yield
$z_E=2-y/3$~\cite{medina,freylong}, an expression which is identical
for a given $y$ to that in the randomly stirred NS Eq.~(\ref{ns}).

\subsubsection{Extinction transition in contact with an Edward-Wilkinson fluctuating
surface}\label{ewrg}

Let us now assume that the percolating agent $\phi$ is coupled to a
fluctuating surface described by the EW equation (\ref{ew}). Thus an
EW surface now forms the environment of the percolating process. On
general symmetry ground the time evolution of the density field
$\phi$ may be written as
\begin{equation}
\frac{\partial\phi}{\partial t} + \lambda_2(\nabla
h)\cdot(\nabla\phi) + \lambda_3\phi \nabla^2 h= D\nabla^2\phi
+\lambda_g\phi -\lambda_d\phi^2 + \sqrt\phi \zeta, \label{kpzphiI}
\end{equation}
where $\lambda_2$ and $\lambda_3$ are coupling constants.
Equation (\ref{kpzphiI}) may be
written as
\begin{equation}
\frac{\partial\phi}{\partial t} + \hat\lambda(\nabla
h)\cdot(\nabla\phi) + \lambda_3 \nabla\cdot(\phi\nabla h)=D\nabla^2
\phi + \lambda_g\phi -\lambda_d\phi^2 + \sqrt\phi\zeta,
\label{kpzphiII}
\end{equation}
where $\hat\lambda=\lambda_2-\lambda_3$. The $\lambda_3$-term in
Eq.~(\ref{kpzphiII}), being a gradient term, is irrelevant in the
long wavelength hydrodynamic limit (which is our region of interest
here). Thus the effective equation for $\phi$ that we are going to
work with is
\begin{equation}
\frac{\partial\phi}{\partial t} + \hat\lambda(\nabla
h)\cdot(\nabla\phi) =D\nabla^2\phi + \lambda_g\phi -\lambda_d\phi^2
+ \sqrt\phi\zeta.  \label{kpzphi}
\end{equation}
Just as in the previous case of a randomly stirred fluid
environment, on grounds of general arguments we expect to find four
different phases characterized by four different fixed points -
Gaussian, DP, LDP and LR. The first two correspond to weak dynamic
scaling where as the last two should display strong dynamic scaling.
The EW Eq.~(\ref{ew}) being linear the corresponding dynamics has a
dynamic exponent $z_E=2$. Thus, strong dynamic scaling for the
density field $\phi$ implies it will display dynamic scaling
characterized by an exponent $z_\phi=2$. Although this corresponds
to its value of the linearized theory [see the mean-field exponents
given by Eq.~(\ref{mean})], it does not really correspond to an AAPT
characterized by the mean-field exponents, since all the other
critical exponents (e.g., $\eta,\nu$ etc) differ from their
mean-field values. Our calculations below confirms this picture.

Using Eqs.~(\ref{kpzphi}) and (\ref{ew}) we can write down the
generating functional for the model given by $ {\mathcal
Z}_{EW}=\int DhD\hat{h} D\phi D\hat{\phi} \exp[S_{EW}]$,
  where $S_{EW}$ is the
action of the model. Function $\hat{h}$ and $\hat{\phi}$ are the
conjugate auxiliary fields which appear due to the averaging over
the noise distributions. Now to simplify the action and the
calculations subsequently we rescale the fields as
$\hat{\phi}\rightarrow i\alpha\hat{\phi}$, $\hat{h}\rightarrow
i\hat{h}$ and $\phi\rightarrow \beta\phi$ with $\alpha\beta=1$. The
parameters are then rescaled as $D_1\alpha^2\beta={D g_2 \over 2}$,
$\lambda_d\alpha\beta^2={D g_1 \over 2}$ and $\lambda_g=\tau D$.
This modified action can be written as
 \bea
  S_{EW}(h, \hat{h},
\phi, \hat{\phi}) &=& D_2\int \frac{d^dk}{(2\pi)^d}\int dt
\hat{h}\hat{h} k^{2-y -d} - \int \frac{d^dk}{(2\pi)^d}\hat{h}\{
\partial_t h + D_h q^2h\} - \int\frac{d^dk}{(2\pi)^d}\hat{\phi}
\{ \partial_t\phi - \hat{\lambda} \sum_{\bf q}h({\bf q})\phi({\bf
k-q}){\bf q}\cdot {\bf (k-q)} \nonumber \\&+& D k^2\phi - \tau D\phi
+ {D \over 2}\sum_{\bf q}[g_1\phi({\bf q})\phi({\bf k-q}) -
g_2\hat{\phi}({\bf q})\phi({\bf k-q})]\}. \label{actew}\eea
 Note that action (\ref{actew}) is not invariant
 under the rapidity transformation, similar to action
 (\ref{action1}) .
As before, we are required to identify all the primitive divergent
one-loop vertex functions. The vertex generating functional
$\Gamma_{EW}$ is defined in the standard way as $\log {\mathcal
Z}_{EW}$. The EW equation (\ref{ew}) being linear, there are no
corrections to the vertex functions $\frac{\delta^2
\Gamma_{EW}}{\delta h \delta \hat h}$ and $\frac{\delta^2
\Gamma_{EW}}{\delta\hat h\delta\hat h}$, where $\Gamma_{EW}$ is the
Legendre transformation of $\log Z_{EW}$. However, there are now
non-zero one-loop fluctuation corrections to $\hat{\lambda}$. Thus,
the vertex functions that are to be renormalized in order to render
the action (\ref{actew}) finite are (i) $\frac{\delta^2
S_{EW}}{\delta\phi\delta\hat\phi}$, (ii) $\frac{\delta^3
S_{EW}}{\delta\hat\phi\delta\phi\delta\phi}$, and (iii)
$\frac{\delta^3 S_{EW}}{\delta\hat\phi\delta\hat\phi\delta\phi}$.
The (bare) coupling constants for the present problem are $u_R=g_1
g_2,\,w=\frac{\lambda^2}{D^3}$. In addition, dimensionless Schimdt
number  $\theta = D_h/D$ appears as a control parameter. Assuming
renormalizability, we introduce renormalization $Z$-factors for each
of the primitive divergent vertex functions: (i) $\frac{\delta^2
\Gamma_{EW}}{\delta \phi\delta\hat\phi}$, (ii)
$\frac{\delta^3\Gamma_{EW}}{\delta \phi\delta\phi\delta\hat\phi}$,
(iii) $\frac{\delta^3\Gamma_{EW}}{\delta\hat\phi\delta\hat\phi
\delta\phi}$, and (iv) $\frac{\delta^3 \Gamma_{EW}}{\delta
h\delta\hat \phi\delta \phi}$. We use a minimal subtraction scheme,
in which diverging parts of the associated one-loop fluctuation
correction integrals are evaluated in inverse power series of
$\epsilon=4-d$ and $y$.

The renormalized fields and the parameters are related to the
corresponding bare quantities through the relation
 \bea \phi=Z^R\phi^R \,\,,\,\,
\hat{\phi}=\hat{Z}^R\hat{\phi}^R,\,\,\, \tau=Z_\tau\tau^R \,\, , \,\, D=Z_DD^R
\,\, , \,\, g_1=Z_{g_1}g_1^R \,\,,\,\,
g_2=Z_{g_2}g_2^R\,\,\,\mbox{and}\,\,\,\hat\lambda=Z_{\hat\lambda}\hat{\lambda^R}
. \label{zparaew} \eea We evaluate the $Z$-factors of the
renormalized action in terms of the coupling constants $u=g_1g_2$
and $\hat w=D_1\hat\lambda^2/D^3$ and rescale these coupling
constants with a factor of $1/16\pi^2$ as we have done before in
Section \ref{rand}. Writing  {\em Schmidt number} $\theta={D_h \over
D}$, the $Z$-factors are written as
 \bea
 Z&=&\hat Z=1+\frac{u}{8}\frac{\mu^{-\epsilon}}{\epsilon},\\
Z_\tau &=& 1-{3u \over 8\epsilon}\mu^{-\epsilon} - {\hat{w}(1-\theta) \over 2\theta(1+\theta)^2y}\mu^{-y}, \\
Z_D &=& 1-{u \over 8\epsilon}\mu^{-\epsilon} + {\hat w (1-\theta) \over 2\theta(1+\theta)^2y}\mu^{-y}, \label{zgammaew}\\
Z_u &=& 1+{3u \over 2\epsilon}\mu^{-\epsilon} - {\hat w (1-\theta) \over \theta(1+\theta)^2y}\mu^{-y}
- {2\hat w \over \theta(1+\theta)y}\mu^{-y}, \\
Z_{\hat\lambda} &=& 1 + {u \over 8\epsilon}\mu^{-\epsilon} +{\hat w
\over 2\theta(1+\theta)^2y}\mu^{-y}. \label{zlambda1} \eea

The $Z$-factor $Z_\theta$ can be easily calculated from
Eq.~(\ref{zgammaew}) as there is no renormalization for $D_h$:
 \bea
  Z_\theta=Z_D^{-1}=1+{u
\over 8\epsilon}\mu^{-\epsilon} - {\hat w(1-\theta) \over
2\theta(1+\theta)^2y}\mu^{-y}.
 \eea
 The renormalized coupling constants are
defined as
 \bea
u^R=uZ_u^{-1}\mu^{-\epsilon} \,\,,\,\, \hat {w}^R=\hat wZ_{\hat
w}^{-1}\mu^{-y}, \,\,,\,\, \theta^R=\theta
Z_{\theta}^{-1},\,\,\,w^R=w Z_w^{-1}\mu^{-y}.\eea The critical
exponents are derived from these flow functions at the fixed points
of the model which are evaluated easily from the beta functions \bea
\beta_u = \mu{\partial \over \partial\mu} u^R \,\,,\,\, \beta_{\hat
w} = \mu{\partial \over \partial\mu} w^R \,\,,\,\, \beta_\theta =
\mu{\partial \over \partial\mu}\theta^R. \eea
 At the RG fixed point then we have
  \bea \beta_{\hat w}= w^R\left(-y + {5u^R \over 8}+{w^R \over
\theta^R(1+\theta^R)^2}
- {3\hat{w}^R(1-\theta^R) \over 2\theta^R(1+\theta^R)^2}\right), \nonumber \\
\beta_u= u^R\left( -\epsilon + {3 \over 2}u^R - {\hat
w^R(1-\theta^R) \over \theta^R(1+\theta^R)^2}
-{2\hat w^R \over \theta^R(1+\theta^R)} \right), \nonumber \\
\beta_\theta = \theta^R \left( {1 \over 8}u^R - {\hat
w^R(1-\theta^R) \over 2\theta^R(1+\theta^R)^2}\right). \eea

Fixed points of the model can be obtained from
$\beta_u,\,\beta_{\hat w}$ and $\beta_\theta$ by setting them all to
zero. Note that the value of $\theta_R$ at the fixed points cannot
exceed $1$. In fact the range is $1\leq\theta_R\leq 0$. The
different fixed point solutions are as follows:
\begin{itemize}
\item Gaussian fixed point: $u^R=0=\hat w^R$, $\theta^R=\theta$ (bare value of the Schmidt number which is any
finite positive number).
\item DP fixed point:$u^R=2\epsilon/3$, $\hat w^R=0$, $\theta^R=0$.
\item Long range fixed point (LR): $u^R=0, \hat w^R=4y,\theta^R=1$.
\item Long range DP fixed point (LDP): $u^R\neq 0,\,\hat w^R\neq
0,\,\theta^R < 1$. We discuss this fixed point in details below.
\end{itemize}
At the LDP fixed point values of $u^R$, $\hat w^R$ and $\theta^R$
satisfy the coupled equations
\begin{eqnarray}
\frac{3}{2}u^R - 2\frac{\hat w^R}{\theta^R (1+\theta^R)}-\hat
w^R\frac{1-\theta^R}{\theta^R (1+\theta^R)^2}&=&\epsilon,\label{ur}\\
\frac{u^R}{2}+2\frac{\hat w^R}{\theta^R (1+\theta^R)^2} - \frac{\hat
w^R
(1-\theta^R)}{\theta^R (1+\theta^R)^2}&=& y,\label{wr}\\
\frac{u^R}{4} - \hat w^R \frac{1-\theta^R}{\theta^R
(1+\theta^R)^2}&=&=0.\label{thetar}
\end{eqnarray}
The solutions are given by
\begin{equation}
u^R=4(2y+\epsilon)/9,\;\;\;\hat w^R=\frac{2}{9}
\frac{(5y-2\epsilon)^2}{y-\epsilon},\;\;\;\theta^R=
3\frac{y-\epsilon}{7y-\epsilon}. \label{ewvalues}
\end{equation}
Fixed point values (\ref{ewvalues}) formally define the LDP fixed
point. Positivity of $\theta_R$ demands that solutions
(\ref{ewvalues}) are meaningful only if either $y>\epsilon$ or
$7y<\epsilon$, such that the ratio $(y-\epsilon)/(7y-\epsilon)$ is
positive. However, when $y<\epsilon$ and $7y>\epsilon$, $\theta$ is
negative and hence solutions (\ref{ewvalues}) are not physical. With
the knowledge of the fixed point values of the coupling constants,
one can now easily write down the corresponding scaling exponents:
At the DP fixed point $({2\epsilon \over 3},0,0)$, the anomalous
dimension $\eta_\phi=\eta_{\hat\phi}$, the dynamic exponent $z$ and
the correlation length exponent $\nu$ are given by
 \bea \eta_\phi=\eta_{\hat{\phi}}={\epsilon \over
12}\,\,,\,\, z_\phi=2-{\epsilon \over 12} \,\,,\,\, {1 \over \nu}= 2
+ {\epsilon \over 4}.
 \eea
  At the LR fixed point $(0,4y,1)$, the same
critical exponents turn out to be \bea
\eta_\phi=\eta_{\hat{\phi}}=0\,\,,\,\, z_\phi=2 \,\,,\,\, {1 \over
\nu}= 2. \eea
 Lastly, at the LDP fixed point we find
\begin{eqnarray}
\eta_\phi &=& \eta_{\hat\phi}=\frac{1}{18}(2y+\epsilon),\\
\nu^{-1}&=&2-\gamma_\tau = 2+ \frac{1}{9}(2y+\epsilon),\\
z_\phi&=&2.
\end{eqnarray}
Thus, clearly both the LDP and LR fixed points correspond to strong
dynamic scaling as the dynamic exponent of the density field $z_\phi=2=z_E$,
the dynamic exponent of the EW surface. Although this is identical to the
mean-field value of $z_\phi$, the scaling behaviors described by the
LR and LDP fixed points do not correspond to mean-field scaling as
can be easily seen from the values of the other exponents.

Finally let us briefly consider linear stability analysis of the
different phases, which can be performed in a standard way analogous
to the previous case of randomly stirred fluid environment. We find,
unsurprisingly, that the Gaussian FP is always unstable for any
$\epsilon >0$ and $y>0$. The DP FP is stable for $\epsilon > 3y$. In
contrast the LR FP is {\em always} unstable, unlike  the case when
the environment is modeled by a randomly stirred fluid. We do not
discuss the stability of the LDP FP due to the associated algebraic
complications. However, the determinant of the stability matrix at
the LDP FP vanish for $y=\epsilon$, suggesting borderline between
stability and instability. Incidentally, the line $y=\epsilon$
demarcates the region of existence of the LDP FP. Further, since the
LDP FP does not exist for $y<\epsilon<7y$ and DP FP is unstable for
$\epsilon <3y$, there is no physically meaningful solution in the
region $y<\epsilon < 3y$, a situation that does not arise when the
environment is a randomly stirred fluid. Fig.~(\ref{ewfig}) below
shows the stable phases of the model with a fluctuating EW surface
as the environment.
\begin{figure}[htb]
\includegraphics[height=5cm]{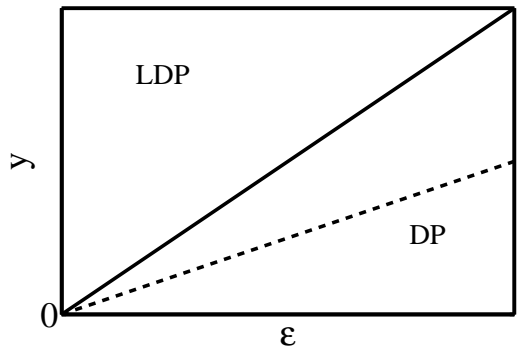}
\caption{Schematic phase diagram depicting the stable phases of the
model with an EW surface as the environment. The continuous line is
given by $y=\epsilon$ such that the LDP phase exists for
$y>\epsilon$. The dashed straight line is given by $y=\epsilon/3$;
the usual DP phase exists for $\epsilon >3y$. There is no
physically meaningful phase in the region between the lines
$y=\epsilon$ and $y=\epsilon/3$.} \label{ewfig}
\end{figure}

\subsubsection{AAPT in contact with a fluctuating
KPZ surface} \label{kpzrg}

We now briefly consider how a fluctuating KPZ surface may affect the
universal properties of extinction transition of a population
density $\phi$ near its threshold. Due to associated algebraic
complications our studies in this section are less extensive.
Nevertheless we are still able to obtain physically interesting
results consistent with the results obtained elsewhere in this
paper. Field $\phi$ follows Eq.~(\ref{kpzphi}), as in the case when
the environment is modeled by a growing EW surface. However, unlike
the case of the environment modeled by the EW Eq.~(\ref{ew}) for the
specific choice $\lambda_1=\lambda_2$, Eqs. (\ref{kpz}) and
(\ref{kpzphiI}) are invariant under the tilt transformation:
\begin{equation}
h\rightarrow h+ {\mathbf\varepsilon}\cdot {\bf
x},\,\,\phi\rightarrow\phi,\,\,\,{\bf x}\rightarrow {\bf x} -
\lambda_1 {\mathbf \varepsilon} t.\label{tilt1}
\end{equation}
As previously, Eq.~(\ref{kpzphiI}) reduces to Eq.~(\ref{kpzphi})
after discarding total derivative terms in the long wavelength
limit. Function $\xi$ is a zero mean Gaussian white noise with a
variance (\ref{vari1}). As before, the multiplicative nature of noise
ensures that the absorbing state $\phi=0$ is a solution of
Eq.~(\ref{kpzphi}). Similar to the case of a randomly stirred fluid
environment, on general physical ground we expect four different DRG
fixed points to exist, corresponding to four distinct phases: (i)
Gaussian, (ii) DP, (iii) LDP, (iv) LR. The Gaussian FP is expected
to be always unstable. The DP FP corresponds to weak dynamic
scaling, where as phases corresponding to the LDP and LR FPs should
display strong dynamic scaling.

For performing DRG calculations, we use a path integral formulation
in terms of the Janssen-De Dominicis generating functional
corresponding to the KPZ Eq.~(\ref{kpz}) together with the Gaussian
long range noise with a variance (\ref{kpznoise}) for correlation
functions as before. Rescaling as for the case with a fluctuating EW
surface as the environment, the action functional now reads [after
dropping total spatial derivative terms, see the discussions
preceding (\ref{actew})]
 \bea
  S_{KPZ}(h,\hat h, \phi, \hat{\phi})
&=& 2D_2\int \frac{d^dk}{(2\pi)^d}\int dt \hat h \hat h k^{2-y -d}
- \int\frac{d^dk}{(2\pi)^d}\int dt\hat h\{ \partial_t h - {\lambda_1 \over 2}
\sum_{\bf q} {\bf q\cdot(k-q)}h({\bf q})h({\bf k-q}) + D_h k^2h\} \nonumber \\
&&- \int\frac{d^dq}{(2\pi)^d}\int dt\hat{\phi}[\partial_t\phi -
\hat\lambda\sum_{{\bf q}}{\bf q\cdot (k-q)} h({\bf q})\phi({\bf
k-q}) + D k^2\phi - D\tau\phi \nonumber \\&+& {D  \over 2}\sum_{\bf
q}[g_1\phi({\bf q})\phi({\bf k-q}) - g_2\hat{\phi}({\bf q})\phi({\bf
k-q})]]. \label{actkpz} \eea Here again $\hat h$ and $\hat \phi$ are
auxiliary (conjugate) fields.
 Like (\ref{action1}), action (\ref{actkpz}) is no longer
 invariant under the rapidity symmetry. Consequently,
 nonlinear coefficients $g_1$ and $g_2$ in general are unequal.
As for the usual DP problem, the system exhibits a continuous phase
transition from active to absorbing states as (renormalized or
effective) $\tau\rightarrow 0$. The associated universal scaling
exponents are formally defined as in (\ref{scaling}) above. At the
mean-field level, the model Eqs.~(\ref{kpz}) and (\ref{kpzphi})
yield the same values for the scaling exponents as in
Sec.~\ref{review}. Just like the model in Sec.~\ref{rand}, nonlinear
couplings $\lambda$ and $Dg_1,\,Dg_2$, together with the expected
large fluctuations near the critical point and multiplicative nature
of the noise with the long-ranged variance (\ref{yakhotnoise})
substantially alter the mean-field values of the exponents, a fact
which we confirm below by our one-loop DRG calculation. Although in
general $g_1\neq g_2$, it is only the product $g_1g_2$ that appears
in the perturbative expansion.

The structure of the perturbation theory and its renormalization is
very similar to those in Sec.~\ref{rand} above. Due to the long
ranged nature of the noise variance (\ref{kpznoise}) there is no
renormalization of $D_1$. As in the previous case in
Sec.~\ref{ewrg}, the role of the coupling constants in the ordinary
perturbation theory in the present model is played by
$u=g_1g_2,\;\;\;w=\frac{1}{D_h^3},\,\,\,\hat
w=\frac{\hat\lambda^2}{D^3}$.
We again consider $\epsilon
>0,\,y>0$, for which
non-trivial critical exponents will ensue.  The vertex generating
functional $\Gamma_{KPZ}$ is defined as the Legendre transformation
of $\log {\mathcal Z}_{KPZ}$. Galilean invariance of the action
functional (\ref{actkpz}) ensures that the three-point vertex
function $\frac{\delta^3 \Gamma_{KPZ}}{\delta \hat h\delta h\delta
h}$ does not renormalize in the hydrodynamic limit. The vertex
functions, which must be renormalized in order to render the present
model renormalized, are (i) $\frac{\delta^2 S_{KPZ}}{\delta h \delta
\hat h}$, (ii) $\frac{\delta^2 S_{KPZ}}{\delta\phi\delta\hat\phi}$,
(iii) $\frac{\delta^3 S_{KPZ}}{\delta\hat\phi\delta\phi\delta\phi}$,
(iv) $\frac{\delta^3
S_{KPZ}}{\delta\hat\phi\delta\hat\phi\delta\phi}$, and
$\frac{\delta^3 S_{KPZ}}{\delta\hat\phi\delta\phi\delta h}$. Using
standard methods as described in the previous Sections we in details
perform one-loop multiplicative renormalization by introduction of
the renormalization $Z$-factors, which render the theory UV finite.
We use dimensional regularization together with minimal expansion to
enumerate the renormalization $Z$-factors. From the renormalization
$Z$-factors, one may then derive the RG flow equation in the usual
way. Finally, the scaling behavior of the correlation or vertex
functions may be extracted by finding their dependence on $\mu$ by
using the RG equation derived below. The renormallization
$Z$-factors for the fields and the parameters are defined as
 \bea \phi=Z\phi^R \,\,, \,\,
\hat{\phi}=\hat{Z}\hat{\phi}^R,\,\,\, \tau=Z_\tau\tau^R \,\, , \,\,
D=Z_D D^R \,\, , \,\, g_1=Z_{g_1}g_1^R, \,\,,\,\, g_2=Z_{g_2}g_2^R,
\,\, \hat\lambda=Z_{\hat\lambda}\hat\lambda^R,\,\, \mbox{and} \,\,
D_h=Z_{D_h} D_h^R. \label{zparab} \eea

We evaluate the $Z$-factors of the renormalized action in terms of
the coupling constants $u=g_1g_2$ and $w=D_2/D_h^3$ and rescale
these coupling constants with a factor of $1/16\pi^2$ as we have
done before in Section \ref{rand2}. The different $Z$-factors are
\bea
 Z&=&\hat Z=1+\frac{u}{8}\frac{\mu^{-\epsilon}}{\epsilon},\\
Z_\tau &=& 1+{3u \over 8\epsilon}\mu^{-\epsilon} + {\hat{w} \over 2\theta (1+\theta)}\frac{1-\theta}{1+\theta}\frac{\mu^{-y}}{y}, \\
Z_D &=& 1-{u \over 8\epsilon}\mu^{-\epsilon} - {\hat{w} \over {2\theta (1+\theta)}}\frac{1-\theta}{1+\theta}\frac{\mu^{-y}}{y}, \label{zgamma}\\
Z_u &=& 1+{3u \over 2\epsilon}\mu^{-\epsilon} - {2\hat{w} \over {\theta(1+\theta)}}\frac{\mu^{-y}}{y}-\frac{3}{2} \frac{\hat w(\theta-1)}{2\theta (1+\theta)^2}\frac{\mu^{-y}{y}}, \\
Z_{\hat\lambda}&=&1+ \frac{u}{8}\frac{\mu^{-\epsilon}} + \frac{\hat
w}{\theta (1+\theta)^2}\frac{\mu^{-y}}{y}-\sqrt w
\sqrt {\hat w} \frac{1+3\theta}{\sqrt\theta (1+\theta)^2},\\
Z_{\hat w}&=&1+\frac{5u}{8}\frac{\mu^{-\epsilon}}{\epsilon} +
\frac{2\hat w}{\theta (1+\theta)^2}\frac{\mu^{-y}}{y} + \frac{3\hat
w}{2\theta (1+\theta)}\frac{\theta -1}{\theta +1}\frac{\mu^{-y}}{y}
- 2\sqrt w\sqrt{\hat w} \frac{1+3\theta}{\sqrt\theta
(1+\theta)^2},\\
 Z_{D_h}
&=& 1- {wb \over y}\mu^{-y},
 \label{znu} \eea
where $b=y/4d$\cite{medina}.
%

The renormalized coupling constants are written as \bea
u^R=uZ_u^{-1}\mu^{-\epsilon} \,\,,\,\, w^R=wZ_w^{-1}\mu^{-y}
\,\,,\,\,
\hat{w}^R=\hat{w}Z_{\tilde{w}}^{-1}\mu^{-y},\,\,\,\theta^R=\theta
Z_\theta^{-1}. \eea The critical exponents are obtained from the
Wilson's flow functions at the fixed points of the model which are
evaluated easily from the zeros of the beta functions \bea \beta_u =
\mu{\partial \over
\partial\mu} u^R \,\,,\,\, \beta_w = \mu{\partial \over \partial\mu}
w^R \,\,,\,\, \beta_{\tilde{w}} = \mu{\partial \over
\partial\mu}\tilde{w}^R,\,\,\,\beta_\theta=\mu\frac{\partial}{\partial \mu}\theta^R. \eea
At the RG fixed point then we have
 \bea
  \beta_{\hat w}&=& \hat w^R\left[-y + \frac{5}{8}u^R + \frac{2\hat w^R}{\theta^R (1+\theta^R)^2}-
\frac{3\hat w^R}{2\theta^R
(1+\theta^R)}\frac{1-\theta^R}{1+\theta^R} -2\sqrt {w^R}\sqrt{\hat
w^R}\frac{1+3\theta^R}{\sqrt{\theta^R}
(1+\theta^R)^2}\right],\nonumber\\
\beta_u&=& u^R\left[ -\epsilon + {3 \over 2}u^R -
\frac{2\tilde{w}^R}{\theta^R(1+\theta^R)}+\frac{\hat w^R}{\theta^R
(1+\theta^R)}\frac{\theta^R-1}{\theta^R+1} \right],\nonumber \\
\beta_{{w}}&=& {w}^R \left( -y + 3bw^R\right),\nonumber \\
 \beta_\theta&=&\theta^R \left[\frac{u^R}{8}-\frac{\hat w^R}{2\theta^R(1+\theta^R)}\frac{1-\theta^R}{1+\theta^R}-bw^R\right]. \eea
It is clear that $w^R=y/3b$ is the only stable fixed point solution
for $w^R$. The other solution $w^R=0$ is unstable always and we
ignore it from our discussions below. While we find the FPs and the
corresponding scaling exponents below, we do not discuss their
linear stability. Although the latter can in principle be done just
as we do for the other models, it is algebraically much more
complicated due to the structure of the fixed point equations. As
before, we expect four different fixed points to exist, similar to
the previous cases. We also expect weak and strong dynamic scaling
in different situations. We find
\begin{itemize}
\item $\,\, u^R=0, \hat{w}^R=0 \,\,\, \mbox{(Gaussian fixed point)}$. The corresponding scaling exponents
are those of the linearized system. \\
\item $\,\, u^R={2 \over 3}\epsilon, \hat{w}^R=0 \,\,\,
\mbox{DP fixed point}$. The exponents are given by those of the DP universality class. \\
\item $\,\, u^R=0; \hat w^R \;\;{\rm and}\;\; \theta^R $ may be solved from the coupled nonlinear equations
\begin{eqnarray}
\frac{\hat w^R}{2\theta^R (1+\theta^R)}
\frac{1-\theta^R}{1+\theta^R} =
-bw^R = y/3,\\
\frac{2\hat w^R}{\theta^R (1+\theta^R)^2} -\frac{3\hat
w^R}{2\theta^R (1+\theta^R)}\frac{1-\theta^R}{1+\theta^R} - 2\sqrt
{w^R} \sqrt {\hat w^R}\frac{1+3\theta^R}{\sqrt \theta^R
(1+\theta^R)^2}=y.
\end{eqnarray}
Explicit solutions of the above equations are a difficult task,
owing to their highly nonlinear nature. However, without their
explicit solutions, one may already obtain the following
information: (i) Since $\hat w^R>0$, one has $\theta^R >1$, thus
$D>D_h$, (ii) Dynamic exponent $z_\phi = 2-\gamma^*_D=2-\frac{\hat
w^R}{2\theta^R (1+\theta^R)}\frac{1-\theta^R}{1+\theta^R}=2-y/3=z_E$
and hence strong dynamic scaling, (iii) correlation length exponent
$\nu$ is given by $\nu^{-1}=2-\gamma_\tau^*=2+y/3$, and (iv) the
anomalous dimension $\eta_\phi^*=\eta_{\hat\phi}^*=u^R/8=0$ at the
LR FP.
\item LDP fixed point: $u^R\neq 0$.  Actual enumeration of the fixed
point values of the coupling constants are very difficult due to the
complicated nonlinear structures of the underlying equations, and
will not be discussed here. Using, however, the approximation
$\theta^R \gg 1$, we obtain $u^R=4(\epsilon +2y/3)/7$. Fixed point
values $\theta^R$ and $\hat w^R$ are to be obtained from the coupled
nonlinear equations
\begin{eqnarray}
\sqrt{\hat w^R}=\theta^R\sqrt{\frac{4y-\epsilon}{21}},\label{thetldp}\\
 \frac{\epsilon+2y/3}{7} +\frac{2\hat w^R}{\theta^R (1+\theta^R)^2}
- \sqrt {w^R} \sqrt{\hat w^R} \frac{1+3\theta^R}{\sqrt{\theta^R}
(1+\theta^R)^2}=0.
\end{eqnarray}
With $w^R=y/3b$ the above two equations may in principle be solved
and solutions be obtained with the overall approximation of large
$\theta^R$. We do not solve these here explicitly. Nevertheless, we
can already extract useful information without having to solve for
the coupling constants explicitly. We find: (i) dynamic exponent
$z_\phi = 2-\gamma^*_D=2-\frac{u^R}{8}-\frac{\hat w^R}{2\theta^R
(1+\theta^R)}\frac{1-\theta^R}{1+\theta^R}=2-y/3=z_E$ and hence
strong dynamic scaling, (ii) correlation length exponent $\nu$ is
given by $\nu^{-1}=2-\gamma_\tau^*=2+3u^R/8 - \hat
w^R(1-\theta^R)/[2\theta^R (1+\theta^R)^2] =2+u^R/4 +bw^R=2+
(\epsilon +2y/3)/7 + y/3$, (iii) anomalous dimension
$\eta_\phi^*=\eta_{\hat\phi}^*=u^R/8=(\epsilon+2y/3)/14$, and (iv)
by using positivity of $\theta^R$ and $\hat \omega^R$, $4y>\epsilon$
from Eq.~(\ref{thetldp}) for physically meaningful solution.
\end{itemize}
Thus,  with a KPZ surface as a fluctuating environment for an
extinction transition that, otherwise (i.e., with a uniform
environment) belongs to the DP universality class, the broad
emerging picture is similar to the other two models of fluctuating
environment considered here before. One generally finds both weak
and strong dynamic scaling in different regions of the phase space
spanned by $\epsilon$ and $y$. The details, including the values of
the scaling exponents, may of course depend upon the actual model of
the fluctuating environment. Lastly, some technical comments
regarding alternatives to the DRG procedure here is in order: As we
commented before, the one-loop DRG procedure for the KPZ
Eq.~(\ref{kpz}) with long ranged noise suffer well-known technical
problems. As an alternative to it, self-consistent mode coupling
method (SCMC) \cite{scmc} and functional renormalization group (FRG)
\cite{frg} have been used to extract large length-scale, long-time
physics of the KPZ Eq. with long ranged noise. The SCMC expansion,
as illustrated in Ref.~\cite{scmc}, yields results which match with
those in Ref.~\cite{medina} at $1d$, where as for $d>1$ the results
of Ref.~\cite{scmc} differ substantially from Ref.~\cite{medina} and
yields much more physically sensible results in the limit of short
ranged noise. Similarly, Ref.~\cite{frg} uses the well-known
Cole-Hopf transformation and applies the FRG (up to two-loop) on the
resulting partition function to obtain scaling exponents for the KPZ
Eq.~(\ref{kpz}) with long-ranged noises. Their results clearly
highlight the short comings of the one-loop DRG procedure. While the
qualitative picture that emerges out of our one-loop DRG
calculations here are expected to remain on general physical
grounds, it will be interesting to calculate the details of the AAPT
transition in contact with a KPZ surface by using the SCMC or FRG
methods as illustrated in Refs.~\cite{scmc,frg}. The EW
Eq.~(\ref{ew}) being linear such technical issues as for the KPZ Eq.
do not arise. Nevertheless, investigation of the associated AAPT
transition (in contact with an EW surface) by using SCMC or FRG
methods would be useful.

\section{Summary and outlook}\label{summary}

This article is a study of how non-trivial fluctuating
spatio-temporal dynamics of the environment may affect the usual
directed percolation process with constant environment. We have
separately considered cases when the environment is a (i) randomly
stirred fluid described by the Navier-Stokes equation with a
long-ranged force, (ii) a fluctuating surface with long-ranged
spatial correlation, described either by the KPZ or the EW equations
driven by a long-ranged noise. Our model systems are {\em
semi-autonomous}, i.e., we ignore feedback due to the percolating
field on the environment. The general picture that emerges out of
our calculations is that depending upon relative values of
$\epsilon=4-d$ and $y$, a parameter that fixes the spatial scaling
of variances of the external forces in the NS, EW or KPZ equations,
one obtains different universal behavior. However, the details of
the ensuing phase diagram in the $\epsilon - y$ plane depend
explicitly on the model used to describe the environment (NS, EW or
KPZ). On general ground we predict possibilities of four different
phases (i) Gaussian, (ii) original DP, (iii) Long-range DP (LDP) and
(iv) Long-range (LR). The Gaussian fixed point is generally
unstable. The rest are model dependent (i.e., depends upon whether
the environment is modeled by the NS, EW or KPZ Eq.). For a randomly
stirred environment described by the NS Eq., we find when $\epsilon
> 4y$, the ensuing universal critical behavior of the AAPT
transition is described by the standard DP universality class
characterized the DP fixed point with the renormalized coupling
constants $u^R=2\epsilon/3$, $\tilde w^R=0$, corresponding to a
dynamic exponent $z_\phi<z_E$, the dynamic exponent of the
environment. Thus one obtains weak dynamic scaling, despite the
nonlinear coupling between $\phi$ and the environment (here:
velocity $\bf v$). In the other regime, i.e., when $\epsilon <4y$,
the DP fixed point gets unstable against perturbations due to the
environmental fluctuations, and the AAPT is described by a set of
critical exponents that depend upon $y$. The corresponding phases
are described by either the LDP fixed point $u^R=4\epsilon/5- 8y/15$
and $\tilde w^R=8y/15-2\epsilon/15$, or by the LR fixed point
$u^R=0$ and $\tilde w^R=4y/9$. In both these regimes
$z_\phi=z_E=2-y/3$, thus describing strong dynamic scaling. We
obtain the relevant scaling exponents in all the phases.
Agreement of our results for the scaling exponents and the phase
diagram with those in Ref.~\cite{antonov1} shows the general
robustness of the perturbation theory, despite our using (slightly)
different choices for the coupling constants. It also shows how such
choices may be exploited to infer the large length-scale, long-time
limit physics of the system in a simple manner. In contrast, when
the environment is a fluctuating surface modeled by the EW Eq., the
DP phase is linearly stable for $\epsilon
> 3y$. We further find that the LR phase is not at all stable for
any $y>0$. Furthermore, the LDP phase does not exist in the range
$y<\epsilon<7y$. Thus, the phase diagram in this case does not have
any physically meaningful phase in the range $y<\epsilon<3y$, unlike
the NS case where the phase diagram is fully spanned by the stable
phases of the system. The LDP phase here corresponds to strong
dynamic scaling and the DP phase weak dynamic scaling. We obtain the
associated scaling exponents as well. For an environment described
by a fluctuating KPZ surface, we  again find the existence of four
different phases similar to the previous cases. We are able to
obtain the scaling exponents in each phase and show that strong
dynamic scaling prevails in the LDP and LR phases, as expected.
However, due to the algebraically complicated nature of the DRG
$\beta$-functions we do not discuss their linear stability here.
Validity of our results are limited by the applicability of one-loop
approximations which suffer from well-known technical difficulties
\cite{jkb} in systems with long range noises. Thus it is important
to verify our results in numerical simulations of the models used
here. Further, it is unclear how multiscaling of the velocity field
given by the randomly stirred NS model or the height field given by
the KPZ equation affect the scaling of the percolating agent $\phi$,
or whether $\phi$ itself will display multiscaling for its higher
order structure functions. Since the EW equation is a linear
equation, it does not show any multiscaling. However still, field
$\phi$ may display multiscaling, similar to the passive scalar
problem of fluid turbulence, where even if the velocity field is
Gaussian distributed (albeit with a long ranged spatial
correlation), the passive scalar density exhibits multiscaling
\cite{passive-scal}. We look forward to numerical solutions of the
continuum model equations in resolving the outstanding theoretical
issues discussed above. In addition, the effects of EW or KPZ
fluctuating surfaces on the DP universality may be studied by
numerical simulations of lattice-gas type models, which may be
constructed by borrowing, e.g., the lattice-gas models used in
Ref.~\cite{lattice} to study the dynamics of a passive scalar on
fluctuating surfaces. Effects of turbulent flow on the DP
universality may in principle be investigated by adopting the
experimental set up of Ref.~\cite{chate} and stirring the system
with long-ranged external forces, or by observing grwoth/decay of a
bacteria colony in a turbulent flow.

Our models and results provide for simple examples of {\em weak
dynamic scaling}, situations which are not very commonly found.
Known examples include the studies in
Refs.~\cite{kardar,jkb-amit,das-basu,uwe1}. Our semi-autonomous
models, where  effects of the local population on the environment is
completely neglected, are certainly a simplification of more
realistic situations where such feedback effects should be present
in general. A non-zero feedback is known to affect the scaling
properties of the NESS in general. For instance, the
scaling/multiscaling properties of the magnetic fields in
three-dimensional Magnetohydrodynamics ($3d$MHD) are vastly
different when the feedback (in the form of Lorentz forces in
$3d$MHD) is present from when it is absent (the passive vector
limit) \cite{mhd}. Given this, it would be interesting see how
feedback due to $\phi$ may alter the emerging scaling behavior
discussed here. When a feedback is present, the overall system is no
longer autonomous, but fully coupled. It would be intriguing to see
if weak dynamic scaling persists even in the fully coupled case. A
truly novel feature of our results is that, in all the three
models considered here regardless of stability issues, the LR and
the LDP FPs correspond to the {\em same} value of the dynamic
exponent $z_\phi$, but {\em different} values for the static scaling
exponents $\eta_\phi,\nu$. This is in contrast with what one finds
in equilibrium critical dynamics. For instance, model A and model B
(in the nomenclature of Ref.~\cite{halp}) display different dynamic
exponent but the same static critical exponents for the second order
phase transition in the $O(N)$ model. Equality of the static
critical exponents for different dynamics (i.e., different dynamic
exponents) is a requirement of thermal equilibrium. Since our work
concerns here models that are driven our of equilibrium, such
considerations do not arise. Numerical solutions of the continuum
stochastic equations of motion or simulations of equivalent lattice
models should be able to verify our results. Beyond immediate
theoretical motivation, our work sheds important light for more
realistic problems, e.g., population dynamics of a bacteria colony
or a biofilm resting over a fluctuating surface or a fluctuating
biomembrane. Further, Our results provide important insight for more
biologically motivated problems, e.g., population dynamics of a
bacteria colony over a fluctuating surface, or in the presence of a
macroscopic order, e.g., in a nematic or polar active fluid
\cite{active}.


\section{Acknowledgement}
One of the authors (AB) wishes to thank  the Max-Planck-Society
(Germany) and Department of Science and Technology (India) for
partial financial support under the Partner Group program (2009).


\begin{thebibliography}{99}
\bibitem{grib1} P. Grassberger and K. Sundermeyer, {\em Phys. Lett. B}, {\bf 77},
220 (1978).
\bibitem{grib2}P. Grassberger and A. de La Torre, {\em Ann. Phys.
(N.Y.)}, {\bf 122}, 373 (1979).
\bibitem{rft1} V.N. Gribov, {\em Zh. Eksp. Teor. Fiz.}, {\bf 53}, 654 (1967)
[{\em Sov. Phys. JETP}, {\bf 26}, 414 (1968)].
\bibitem{rft2}V.N. Gribov and A.A. Migdal, {\em Zh. Eksp. Teor. Fiz.}, {\bf 55},
1498 (1968) [{\em Sov. Phys. JETP}, {\bf 28}, 784 (1969)].
\bibitem{rft3} O. Mollison, {\em J. R. Stat. Soc. B (Methodol.)}, {\bf 39}, 283 (1977).
\bibitem{popdyn2} H.K. Janssen, {\em J. Stat. Phys.}, {\bf 103}, 801 (2001).
\bibitem{popdyn3} P. Grassberger, {\em  Z. Phys. B}, {\bf 47}, 365 (1982).
\bibitem{hinrev} A comprehensive recent overview over directed percolation
is available in H. Hinrichsen, {\em Adv. Phys.}, {\bf 49}, 815
(2001); see also  M. Henkel, H. Hinrichsen, and S. L\"ubeck,
Non-Equilibrium Phase Transitions, Berlin: Springer (2008).
\bibitem{uwe-hans-review} For a recent review on the field theory approach to percolation
processes, see H.K. Janssen and U.C. T\"auber, {\em Ann. Phys.
(N.Y.)}, {\bf 315}, 147 (2005).
\bibitem{marro}J. Marro and R. Dickman, Nonequilibrium phase transitions in lattice models (Cambridge
University Press, Cambridge, 1999).
\bibitem{chate} K. A. Takeuchi, M. Kuroda, H. Chat\'e and M. Sano, {\em Phys. Rev. Lett.}, {\bf 99}, 234503 (2007).
\bibitem{dpuniv} H.K. Janssen, {\em Z. Phys.: Cond. Mat. B}, {\bf 42}, 151
(1981).
\bibitem{grass20} P. Grassberger, in {\em Fractals in Physics}, edited by
L. Pietronero and E. Tosatti (Elsevier, 1986).
\bibitem{hin-how} H. Hinrichsen and M. Howard, {\em Eur. Phys. J. B}, {\bf 7}, 635
(1999).
\bibitem{field1} H.K. Janssen, K. Oerding, F. van Wijland, and H.J. Hilhorst,
{\em Eur. Phys. J. B}, {\bf 7}, 137 (1999).
\bibitem{field2} H. K. Janssen and P. Stenull, {\em Phys. Rev. E},
{\bf 78}, 061117 (2008).
\bibitem{antonov} N. V. Antonov, V. I. Iglovikov and A. S. Kapustin,
{\em J. Phys. A: Math. Theor.}, {\bf 42} 135001 (2008).
\bibitem{antonov1} N. V. Antonov, A. S. Kapustin and A. V.
Malyshev, {\em Theor. Math. Phys.} {\bf 169}, 1470 (2011).
\bibitem{landau} L. D. Landau and E. M. Lifshitz, {\em Fluid
Mechanics}, Butterworth-Heinemann (1987).
\bibitem{fns} D. Forster, D. R. Nelson and M. J. Stephen, {\em Phys. Rev A}, {\bf 16}, 732 (1977).
\bibitem{yakhot} V. Yakhot and S. A. Orszag, {\em J. Sci. Comp.}, {\bf 1}, 3 (1986).
\bibitem{foot} In fact, for long range noise with variances
(\ref{yakhotnoise}) the velocity field displays what is known as
{\em multiscaling} in literature. See, e.g., Ref.~\cite{rfnse} for
results and discussions. This is outside the scope of the present
discussion.
\bibitem{jkb} J.K. Bhattacharjee, {\em J. Phys. A}, {\bf 31}, L93 (1998).
\bibitem{ronis} D. Ronis, {\em Phys. Rev. A}, {\bf 36}, 3322 (1987).
\bibitem{bausch} R. Bausch, H.K. Janssen and H. Wagner, {\em Z. Phys. B: Cond. Mat.}, {\bf 24}, 113 (1976).
\bibitem{uwe-book}  Uwe T\"auber, {\em Lect. Notes Phys.}, {\bf 716}, 295
(2007).
\bibitem{foot2} Equivalently, the Galilean
invariance ensures that the combination of terms $\hat\phi
\frac{\partial\phi}{\partial t} + \hat\phi {\bf v}\cdot\nabla\phi$
enter in the renormalized action only in the form of the Lagrangian
derivative $\hat\phi\nabla_t\phi$, where $\nabla_t\equiv \partial_t
+ {\bf v}\cdot\nabla$. Similar considerations apply to $\hat
v_i(\partial_t v_i + {\bf v}\cdot \nabla v_i)$.
\bibitem{barabasi-book} A.-L. Barabasi and H. E. Stanley, {\em Fractal Concepts in Surface
Growth}, Cambridge University Press, Cambridge (1995).
\bibitem{natterman} Lei-Han Tang, T. Nattermann, and B. M. Forrest, {\em Phys. Rev. Lett.}, {\bf 65}, 2422 (1990).
\bibitem{kpzref} M. Kardar, G. Parisi and Y.-C Zhang, {\em Phys. Rev. Lett.}, {\bf 57}, 1810 (1986).
\bibitem{medina} E. Medina, T. Hwa, M. Kardar and Y.-C Zhang, {\em Phys Rev. A}, {\bf 39}, 3053 (1989).
\bibitem{freylong} H.K. Janssen, U.C. T\"auber and E. Frey, {\em Eur. Phys. J. B}, {\bf 9}, 491 (1999).
\bibitem{scmc} E. Katzav and M. Schwartz, {\em Phys. Rev. E}, {\bf
60}, 5677 (1999).
\bibitem{frg} A. A. Fedorenko, {\em Phys. Rev. B}, {\bf 77}, 094203
(2008).
\bibitem{passive-scal}G. Falkovich, K. Gawe,dzki, and M. Vergassola,
{\em Rev. Mod. Phys.}, {\bf 73}, 913 (2001).
\bibitem{lattice} See, e.g., S. Chatterjee and M. Barma,
cond-mat/0509665; A. Nagar, S. N. Majumdar and M. Barma,
cond-mat/0510395, and references therein.
\bibitem{kardar} A. L. Barabasi, {\em Phys. Rev. A}, {\bf 46}, R2977 (1992).
\bibitem{jkb-amit} A. Kr. Chattopadhyay, A. Basu, and J. K. Bhattacharjee,
{\em Phys. Rev. E}, {\bf 61}, 2086 (2000).
\bibitem{das-basu} D. Das, A. Basu, M. Barma, and S. Ramaswamy,
{\em Phys. Rev. E}, {\bf 64}, 021402 (2001)
\bibitem{uwe1} V. K. Akkieni and U. C. T\"auber, {\em Phys. Rev. E},
{\bf 69}, 036113 (2004).
\bibitem{mhd} A. Basu, unpublished.
\bibitem{rfnse} A. Sain, Manu and R. Pandit, {\em Phys. Rev. Lett.}, {\bf 81}, 4377 (1998).
\bibitem{halp} P.C. Hohenberg and B.I. Halperin, {\em Rev. Mod. Phys.} {\bf
49}, 435 (1977).
\bibitem{active}  J.-F. Joanny, J. Prost, in {\em Biological Physics,
Poincar\'e Seminar 2009}, edited by B. Duplantier, V. Rivasseau
(Springer, 2009) pp. 1-32;  S. Ramaswamy, {\em Annu. Rev. Cond.
Matt. Phys.}, {\bf 1}, 323 (2010).
\end{thebibliography}
\end{document}